# Artificial Intelligence and Job Design: A Dynamic Framework

Amin Shirangi*, Todd L. Pittinsky

Department of Technology, AI and Society
Stony Brook University, Stony Brook, NY, USA

## ABSTRACT

This study aims to deepen the understanding of artificial intelligence's (AI) impact on work by introducing and describing the AI-Job Dynamics Framework. Drawing from sociotechnical systems theory, complexity theory, and organizational learning theories, we conceptualize AI integration as a dynamic, non-linear process with three key phases. The framework, supported by theoretical propositions, identifies how this process affects organizational decision-making structures and key job characteristics over time, emphasizing the balance between centralization and decentralization of decision-making power. AI integration initially reduces job autonomy but may lead to greater autonomy in later phases as AI becomes more embedded. Throughout this process, AI reshapes skill requirements from routine to complex cognitive and adaptive skills. In the advanced phase of AI integration, we introduce the alternative paths and varied trajectories through which AI, organizational structures, and individual work experiences evolve. We conclude by discussing the implications of this framework for future research on AI and job design.

## INTRODUCTION

Artificial intelligence (AI) is rapidly reshaping the landscape of work, challenging long-held assumptions about organizational structures, managerial practices, and human capability vis-a-vis AI in non-routine cognitive tasks (Autor et al., 2003). The discourse on AI's impact on work reflects predictions of widespread job displacement (e.g., Acemoglu & Restrepo, 2018; Frey & Osborne, 2017; Korinek & Stiglitz, 2019) and visions of human-AI augmentation (e.g., Agrawal et al., 2023; Autor, 2024; Brynjolfsson, 2023), with these effects potentially co-occurring across sectors. However, many perspectives examine AI at a specific time, presenting its impact as a static, linear progression (e.g., Bughin et al., 2018; Frey & Osborne, 2017). This approach overlooks the nuanced and potentially divergent nature of AI integration, thereby failing to capture the complex, dynamic interplay between AI systems, employees, and organizational structures over time.

This study aims to enhance our understanding of AI's impact on work by introducing the AI-Job Dynamics Framework and presenting a series of theoretically grounded propositions. The framework conceptualizes AI integration as a dynamic, non-linear process with three key phases. The study addresses two interrelated research questions: First, how might the progression of AI integration over time influence the dynamics of decision-making processes in organizations? Second, how might these potential shifts in decision-making structures affect key job design characteristics (Hackman & Oldham, 1976), particularly job autonomy and skill variety?

* Contact: Amin Shirangi (amin.shirangi@stonybrook.edu)

Our interpretation builds on existing literature by considering the non-linear nature of AI integration and its potential impacts on job design. Through a process we identify as AI endogenization, an AI system becomes deeply integrated into organizational processes and culture. This integration can lead to alternative paths, suggesting that different outcomes may emerge as organizations deeply integrate AI. This framework illustrates the diverse ways AI can reshape organizational structures and job design, contributing to ongoing scholarly research on the future of work in AI-augmented environments.

Our analysis of job autonomy and skill variety is grounded in the Job Characteristics Model (JCM) (Hackman & Oldham, 1976), a cornerstone of job design theory. We focus on these characteristics (i.e., job autonomy and skill variety) due to their importance and susceptibility to change in the AI era. AI can both augment and constrain decision-making authority (affecting autonomy) and significantly impact skill requirements. These elements are particularly relevant given their historical trends of increasing over time (Wegman et al., 2018), the concept of skill-biased technological change (SBTC) (e.g., Acemoglu, 1998; Autor et al., 2003; Berman et al., 1998), and AI's potential to reshape organizational decision-making processes.

To advance a better understanding of AI's impact on work, we synthesize insights from three core organizational theories: sociotechnical systems theory (Trist, 1981), complexity theory (Anderson, 1999), and organizational learning theory (e.g., Argyris & Schön, 1978; Levitt & March, 1988). These theories serve as a foundation for conceptualizing the dynamic interplay between AI technologies, job characteristics (specifically job autonomy and skill variety), and organizational structures. Sociotechnical systems theory allows us to examine the interdependencies between AI technologies and social structures within organizations. Complexity theory provides a lens to understand the non-linear, emergent behaviors that may arise from AI integration. Organizational learning theory helps us explore how organizations adapt and evolve their practices as they integrate AI technologies.

We begin by briefly outlining the proposed framework and introducing its components, theoretical underpinnings, and empirical support. We then discuss current limitations in AI and job design research, highlighting the areas the framework seeks to address. Next, we provide a background on the key job characteristics—job autonomy and skill variety—including their definitions and historical trends. This is followed by a detailed exploration of the framework's components, including various AI integration phases and their potential implications for these job characteristics. We conclude with a discussion of the framework's theoretical implications and suggest directions for future research.

## FRAMEWORK OVERVIEW

### Introduction to the Framework

The framework focuses on two key work characteristics: job autonomy and skill variety. The model shows how these two characteristics evolve and interact with AI integration over time. Job autonomy, defined as individuals' degree of control and discretion over their work (Hackman & Oldham, 1976), analyzes employee decision-making power and discretion variations as AI systems are integrated into workplace processes. This component of the model illustrates how AI can both enhance and constrain job autonomy, for a more nuanced understanding of power distribution in AI-augmented workplaces. Skill variety, traditionally defined as the range of different activities and abilities required in a job (Hackman & Oldham, 1976; Morgeson & Humphrey, 2006), examines the changing landscape of skills required in AI-integrated work environments. This component addresses the paradoxical nature of AI's impact on skill requirements, where specific skills may become obsolete while new ones can emerge, resulting in a continuously shifting skill ecosystem.

### AI Integration Phases

Central to the AI-Job Dynamics Framework is the concept of AI integration phases, which we conceptualize as a continuous spectrum reflecting the evolving sophistication and strategic utilization of AI within organizations. This spectrum represents a fluid progression where AI's role shifts from an exogenous force to an increasingly endogenous element of organizational processes. We identify three primary phases along this continuum, each characterized by distinct patterns of AI utilization and organizational adaptation.

***Basic automation.*** The first phase involves applying AI to routine, easily codifiable tasks. In this initial phase, organizations often maintain existing decentralized decision-making structures, with AI serving primarily as an exogenous force to enhance efficiency without fundamentally altering organizational dynamics. Recent empirical evidence suggests that most organizations in the United States remain in this phase, with only 5.8% of US firms adopting AI technologies to produce goods or services. However, adoption rates are significantly higher among very large firms (McElheran et al., 2024).

***AI-driven centralization.*** The second phase is characterized by more advanced AI capabilities that lead organizations to consolidate data and centralize decision-making processes. At this phase, AI systems assume more complex tasks and decision-making roles, resulting in a shift toward centralized structures where decision-making authority becomes increasingly concentrated at higher organizational levels.

***Alternative paths.*** As organizations progress beyond initial AI integration phases, they enter a process we term "AI endogenization." This represents a deep integration of AI into organizational structures and practices, where AI transitions from an external tool to an integral part of the organization's ecosystem. Drawing on complexity theory (Anderson, 1999), we view this as a period of coevolution between human agents and AI systems, which we explore in greater depth later in the paper.

This co-evolutionary process can result in divergent organizational paths. We identify two primary alternative paths: *sustained centralization* and *AI-augmented re-decentralization*. In sustained centralization, organizations solidify their centralized AI-driven decision-making structures, with AI remaining a tool primarily controlled by upper management. Conversely, AI-augmented re-decentralization involves leveraging AI to enable more localized decision-making, often accompanied by widespread reskilling of employees and technologies like edge computing (Cao et al., 2020).

This conceptualization recognizes the complex and non-linear nature of AI's impact on organizational structures. The outcomes of AI endogenization are neither mutually exclusive nor permanent; organizations may cycle between different structural configurations as they navigate this process. Many factors, including organizational culture, industry dynamics, competitive pressures, strategic choices, and human capital, influence an organization's specific trajectory. Notably, both paths present opportunities for human-AI coexistence and complementarity, though the nature and extent of this interaction may vary significantly between scenarios.

**Impact on Job Design Elements**

The framework highlights dynamic shifts in job autonomy and skill variety as organizations progress through AI integration phases. These shifts create a paradoxical tension between automation and augmentation (Agrawal et al., 2023; Raisch & Krakowski, 2021), persisting across time and organizational levels. Job autonomy may initially decrease due to AI-driven centralization but could also increase in later phases as organizations learn to leverage AI to augment human decision-making. This non-linear progression reflects the complex interplay between AI capabilities and organizational learning processes.

The framework also broadens the traditional understanding of skill variety to consider AI's multifaceted impact. Three categories of skills evolve differently through the AI integration phases: routine, complex cognitive, and adaptive skills. Routine skills, which involve standardized, rule-based tasks that follow predetermined procedures (Autor et al., 2003; Frey & Osborne, 2017), will likely decrease as AI systems take over codified tasks. Complex cognitive skills, defined as interconnected mental abilities that allow individuals to analyze and categorize novel problems while applying existing knowledge to unfamiliar situations (Kraft, 2019; van Merriënboer, 1997), may increase, particularly those related to AI interpretation and strategic implementation. Adaptive skills, which encompass the ability to modify behavior and work approaches to respond to novel situations effectively and changing work conditions (Cronshaw & Jethmalani, 2005; Pulakos et al., 2000), will likely gain importance as workers continually adjust to AI-augmented environments.

The demand for these skill types fluctuates as AI systems are integrated, highlighting the simultaneous deskilling and upskilling effects often associated with technological change. This nuanced view of skill variety provides a more comprehensive understanding of AI's impact on job design.

**Theoretical Foundations**

The AI-Job Dynamics Framework is grounded in three core theoretical perspectives. First, we draw on sociotechnical systems theory (Trist, 1981), emphasizing the interdependence between social and technical systems within organizations. We apply its principles to understand the dynamic interplay between AI technologies (as part of the technical system), job characteristics, and human elements (the social system).

Second, complexity theory (Anderson, 1999) captures the non-linear and emergent behaviors arising from complex systems' interactions. In the context of AI integration, it acknowledges that small changes in AI integration can lead to disproportionate and unpredictable shifts in job design and organizational structure. Two key concepts are particularly relevant for us: emergence (Goldstein, 1999) and non-linearity (Casti, 1994). Emergence in the framework is evident in how the interplay between AI integration, job autonomy, and skill variety produces outcomes that cannot be predicted solely from individual components. Non-linearity is highlighted in how small changes in AI integration phases can lead to disproportionate shifts in job characteristics.

Third, organizational learning theory (e.g., Argyris & Schön, 1978; Levitt & March, 1988) explores how organizations create, retain, and transfer knowledge. This perspective is crucial for understanding how learning processes shape the alternative paths of AI integration and its subsequent impacts on job design. Organizational learning plays a pivotal role in mediating the impact of AI integration on job characteristics. In this context, the concepts of exploratory and exploitative learning (March, 1991) are particularly relevant to understanding how organizations adapt to AI technologies. Exploratory learning in AI integration involves holistic experimentation with intelligent machines and searching for new possibilities, such as reimagining decision-making processes, reconceptualizing decision delegation, and identifying emerging skill requirements in AI-augmented environments. Simultaneously, exploitative learning focuses on refining existing AI-integrated processes. By combining these two approaches—exploring new AI possibilities while adjusting current practices—the organization continually deepens its understanding of AI integration. This ongoing learning process shapes decisions about organizational structure, job autonomy, and the skills required.

Synthesizing these theoretical perspectives, the framework emphasizes the reciprocal causality between its components and their technological context. This bidirectional relationship, rooted in complexity theory (Anderson, 1999) and organizational learning (e.g., Argyris & Schön, 1978; Levitt & March, 1988), can lead to unexpected outcomes and tipping points in AI integration. For instance, as organizations transition from basic automation to AI-driven centralization, decreases in job autonomy might trigger increases in workers' adaptive and complex cognitive skills. This shift in skill composition could accelerate progress toward AI endogenization, potentially leading to AI-augmented re-decentralization, where decision-making becomes distributed yet AI-enhanced. This dynamic interplay underscores the non-linear nature of AI integration and its impacts on job characteristics. Two overarching propositions capture these nuanced relationships and form the essence of the framework:

*Proposition 1: As organizations progress through different AI integration phases, the relationship between AI integration and job characteristics (job autonomy and skill variety) will be non-linear and multi-directional rather than following a uniform, unidirectional trajectory.*

*Proposition 2: The impact of AI integration and endogenization on job autonomy and skill variety will be mediated by organizational learning processes. The nature and extent of this mediation will vary across different AI integration phases and potentially influence the organizational outcomes of the AI endogenization process.*

## CURRENT LIMITATIONS IN AI AND JOB DESIGN RESEARCH

The framework addresses several limitations in the research on job design and AI.

First, many studies rely on cross-sectional research designs, which fail to account for the dynamic nature of AI-driven changes over time (Parker et al., 2017). While some recent studies have attempted to address this methodological limitation (Arntz et al., 2016; Holm & Lorenz, 2022), gaps remain. For instance, Holm and Lorenz (2022) investigated how AI use affects skill requirements across different job skill levels using data collected at a single point in time. Such approaches cannot fully capture how AI's impact on job characteristics evolves over time and across various organizational contexts. This methodological constraint limits our ability to theorize about the temporal progression of AI integration and its long-term effects on organizations.

Second, the tendency to treat AI as an exogenous force (Frey & Osborne, 2017) overlooks the potential for reciprocal influences between job design and AI development. This oversight fails to consider how changes in job characteristics might, in turn, shape AI integration and development, creating a feedback loop. For example, Davenport and Kirby's (2016) study of cognitive technologies and augmentation (i.e., human-AI collaboration) does not fully consider how workers' adaptations to AI might, in turn, influence subsequent AI development and deployment.

Third, existing research often assumes a linear progression of AI's impact on organizations (e.g., Arntz et al., 2016; Bughin et al., 2018; Frey & Osborne, 2017). Historical evidence already challenges such linear assumptions. On a broader historical scale, Beaudry et al., (2016, p.227) documented a "great reversal" in the demand for cognitive skills following the IT revolution, with strong growth in demand from 1980 to 2000, followed by an unexpected decline. This pattern illustrates a complex relationship between technology and skill demands.

Contemporary analyses often continue this linear thinking. For example, McKinsey's analysis reflects this linear thinking in two ways. First, it assumes a simple one-to-one relationship where half of the jobs lost to automation are directly replaced by new jobs within the same sector. Second, their skills taxonomy oversimplifies by categorizing complex work activities under single skills - for instance, reducing business correspondence to merely

“advanced literacy,” ignoring the multiple skills like strategic thinking, interpersonal communication, and contextual judgment that such work typically requires (Bughin et al., 2018, p.70). This static approach overlooks how skills combine in unexpected ways and how organizations restructure work during AI integration.

Recent empirical evidence further suggests that AI’s impact follows more complex patterns than linear models would predict. Jia et al., (2024), studying AI-human collaboration in telemarketing, found that AI assistance produced markedly different outcomes across skill levels - significantly enhancing creativity and performance for high-skilled workers while yielding limited benefits for others.

Addressing these limitations requires a theoretical framework that captures the dynamic nature of AI integration and its effects on job design. The framework aims to fill this gap by examining how job autonomy and skill variety coevolve in different phases of AI integration. By conceptualizing AI’s impact as a dynamic and ongoing process rather than a static and fixed one, the approach provides a more nuanced understanding of the interplay between technological advancement and job design in the context of AI implementation.

The next section discusses job autonomy and skill variety within organizational settings. This understanding is essential for understanding the key components of the framework and how they interact during AI implementation.

## JOB AUTONOMY AND SKILL VARIETY

This section provides a comprehensive background on job autonomy and skill variety, two key job characteristics that play significant roles in the AI-Job Dynamics Framework. Understanding their historical evolution and theoretical foundations is essential to understanding how they may be impacted by AI integration.

Job autonomy, one of the most investigated job characteristics (Morgeson & Campion, 2013), has traditionally been conceptualized along three key dimensions (Hackman & Oldham, 1976): work scheduling, decision-making, and work methods. Work scheduling autonomy allows employees to choose their work timings, providing flexibility to manage their work-life balance. Decision-making autonomy enables employees to make significant choices about their work processes and outcomes. Lastly, work methods autonomy involves the authority to determine how tasks are completed, enabling employees to select the techniques and procedures they find most effective (Humphrey et al., 2007; Morgeson & Campion, 2013). Decision-making autonomy is particularly influential among job autonomy dimensions. It is closely linked to increased motivation, job satisfaction, and performance (Morgeson & Humphrey, 2006).

### Theoretical Foundations of Job Autonomy

Two major theoretical frameworks in organizational behavior have significantly shaped our understanding of job autonomy: self-determination theory (SDT) and JCM. SDT identifies autonomy as one of three essential psychological needs, alongside competence and relatedness, that contribute to greater self-motivation and well-being (Deci et al., 1989; Deci & Ryan, 2000). In SDT, autonomy refers to acting with a sense of volition and experiencing personal choice (Gagné & Deci, 2005). Autonomy involves individuals feeling that their actions are their own, reflecting their personal values and interests rather than being imposed by external factors. The JCM, developed by Hackman and Oldham (1976), also emphasizes job autonomy, defining it as the freedom and discretion individuals have in their work. Unlike SDT, JCM focuses primarily on external freedoms within the workplace rather than internalized volition (Parker, 2014).

While SDT and JCM have been influential in understanding job autonomy, they have limitations, particularly in the context of AI integration given AI's potential to dramatically augment and constrain human agency. The JCM, for instance, has been challenged for its limitations in capturing the full complexity of modern work environments (e.g., Bakker & Demerouti, 2007; Parker et al., 2001). More importantly for our focus, neither theory fully accounts for how technological mediation, particularly AI, might fundamentally alter the nature of autonomy in work settings. This points to the need for a new framework to capture the complex interplay between AI and job autonomy.

**Historical Trend Toward Increased Job Autonomy**

Initially, Frederick Taylor's "Scientific Management" (1911) advocated for centralized decision-making and task simplification, effectively reducing worker autonomy. Braverman (1998) argued that Taylor's approach was fundamentally about control rather than efficiency. Interestingly, some aspects of AI integration, such as algorithmic management (Jarrahi et al., 2021; Parker & Grote, 2022), resemble Taylor's principles of standardization and control (Kellogg et al., 2020), potentially signaling a new phase in the evolution of job autonomy.

Mid-20th century theorists challenged this centralized model. Herbert Simon's concept of bounded rationality (Simon, 1957) highlighted the cognitive limitations of decision-makers, while Friedrich Hayek's emphasis on the decentralized nature of knowledge (Hayek, 1945) proposed that distributing decision-making could more effectively leverage collective intelligence and local expertise. These theories laid the groundwork for a shift toward increased worker autonomy, a trend AI may reinforce or disrupt. As organizations became more complex, a trend toward decentralization emerged. Decision-making authority was increasingly delegated to lower organizational levels, enhancing efficiency and responsiveness to local challenges (Bloom et al., 2010). This shift was driven by advancements in information technology, increased human capital, and intensified market competition (e.g., Bresnahan et al., 2002; Caroli & Van Reenen, 2001; Malone, 1997). In this regard, numerous studies confirm a positive trend in job autonomy over recent decades (Kubicek et al., 2017; Wegman et al., 2018).

Skill variety refers to the extent to which a job requires an employee to use a range of skills, abilities, and talents to complete their work (Hackman & Oldham, 1976). It is one of the five core job characteristics in Hackman and Oldham's JCM.

**Theoretical Foundation of Skill Variety**

Skill variety is grounded in theories of job design and human motivation. The JCM posits that skill variety, task identity, and significance contribute to the meaningfulness of work (Hackman & Oldham, 1976). High skill variety implies that a job involves diverse activities, challenging employees to develop and utilize multiple skills. Human capital theory, while not directly focused on job design, supports the broader importance of skill diversity by suggesting that investments in various skills enhance worker productivity and adaptability (Becker, 2009). This principle from human capital theory aligns with and supports the concept of skill variety in job design.

However, these theoretical perspectives, developed in a pre-AI context, may require reconsideration as AI technologies increasingly permeate the workplace. This tension between traditional understandings of skill variety and the emerging realities of AI-integrated workplaces informs our subsequent analysis.

**Historical Trend Toward the Increase in Skill Variety**

The late 20th and early 21st centuries witnessed a significant shift toward increased skill variety in the workforce, driven by the interplay of organizational restructuring and technological advancements. This transformation was characterized by organizations moving from traditional bureaucratic structures to more agile forms, necessitating a multifaceted workforce capable of handling increased responsibilities and uncertainty.

The emergence of the knowledge economy further accelerated this trend. In a knowledge economy, there is greater reliance on intellectual capabilities than physical inputs or natural resources (Powell & Snellman, 2004). This fundamental change prompted organizations to adapt their structures and practices radically. Key organizational changes included decentralization of hierarchies (Malone, 1997), increased use of teamwork (Osterman, 1994), and job rotation (Ichniowski & Shaw, 2003). These changes, known collectively as skill-biased organizational change (SBOC) (Caroli & Van Reenen, 2001), led to an upskilling of the workforce and a demand for a wider variety of skills (Piva et al., 2006).

The rapid adoption of computer technology played a crucial role in amplifying the trend toward increased skill variety. Autor et al. (2003) proposed a framework that conceptualizes work as a series of tasks, classifying them into routine (cognitive and manual) and non-routine (analytical, interactive, and manual) categories. This framework helps explain how technological change has affected skill requirements. Routine tasks became more automated, reducing the need for certain skills, while non-routine tasks, particularly analytical and interactive ones, were complemented by technology, increasing demand for more complex skills. This aligns with the concept of skill-biased technological change (SBTC) discussed earlier. SBTC theory (for reviews, see Acemoglu, 1998; Katz & Murphy, 1992; Morris & Western, 1999), provides a framework for understanding how technological advancements tend to favor skilled labor. It explains the observed trend of increasing demand for skilled workers alongside potential decreases in demand for less-skilled labor as new technologies are adopted.

As a result of these changes, the workplace has seen an increased demand for various types of skills. Cognitive skills such as adaptability, creative problem-solving, critical thinking, and commitment to continuous learning have become increasingly important (Murnane et al., 1995). Soft skills, including interpersonal interactions, teamwork, and emotional intelligence, have grown in importance, driven by team-based structures and the rise of service jobs (Fried et al., 2009; Grant & Parker, 2009). Operational skills like time management and multitasking abilities have also become crucial (Claessens et al., 2007). Supporting this trend, data show a rapid increase in both the supply of education and the returns to more skilled workers (Acemoglu, 2002), likely in response to the growing skill demands of the modern workplace.

These developments ushered in a modern work environment that necessitates a broader range of skills in its workforce. Understanding this historical trend is crucial for anticipating future workforce changes. As AI technologies continue to evolve, they are likely to reshape skill requirements further, potentially accelerating or altering the trajectory observed in recent decades. The framework will examine how AI might influence the nature and variety of skills demanded in the workplace, considering both the potential for skill enhancement and the risk of skill obsolescence.

## FRAMEWORK PHASES AND EFFECTS

In this section, we explore each AI integration phase, considering their potential impacts on organizational decision-making processes, job autonomy, and skill variety. This analysis addresses our two central research questions: how AI integration influences decision-making dynamics over time and how these shifts affect key job design characteristics. We synthesize insights from sociotechnical systems, complexity, and organizational learning theories to uncover the underlying mechanisms driving these dynamics, highlighting theoretical tensions and paradoxes that arise in each phase. The analysis begins with the basic automation phase, where organizations typically initiate their AI journey.

### Basic Automation

Organizations have experienced a trend toward decentralization over recent decades. Within this context, the framework suggests that organizations typically begin their AI journey in a phase characterized by basic automation. Figure 1 depicts this initial phase with high job autonomy and a balanced distribution of skills. In this initial phase, existing decentralized structures largely remain intact. AI systems primarily act as an exogenous force to automate routine, codifiable, and structured tasks without significantly challenging organizational decision-making structures.

*Proposition 3: In the basic automation phase of AI integration, the relationship between routine task automation and (a) employee job autonomy in non-routine tasks, (b) the variety of non-routine skills required, and (c) the variety of routine skills required will be non-linear and characterized by dynamic adaptations.*

***Impact on job autonomy.*** During this initial phase of AI integration, the impact on job autonomy is generally subtle and positive. Organizational decision-making structures largely remain unchanged, with employees typically maintaining their existing levels of autonomy. As basic AI systems begin to handle routine, easily codifiable tasks, employees may have slightly more time and cognitive resources to dedicate to non-routine activities. This modest shift could allow workers to focus more on tasks requiring human judgment, potentially leading to minor enhancements in perceived job autonomy. However, it is essential to note that these changes are typically incremental rather than transformative at this phase.

***Impact on skill variety.*** The skill set required during this phase remains diverse, including routine, cognitive, and adaptive tasks. Autor et al. (2003) provide a foundational framework for understanding this phase, distinguishing between routine and non-routine tasks in the context of computerization. While their work focuses on computers rather than AI specifically, it offers valuable insights that can be extended to early AI integration. In this basic automation phase, initial AI implementations, like early computerization, often target routine cognitive and manual tasks that follow explicit rules. This may leave more complex, non-routine activities to human workers. As a result, there might be a subtle shift in the skill variety demanded from workers, with a slight increase in emphasis on non-routine cognitive and interpersonal skills.

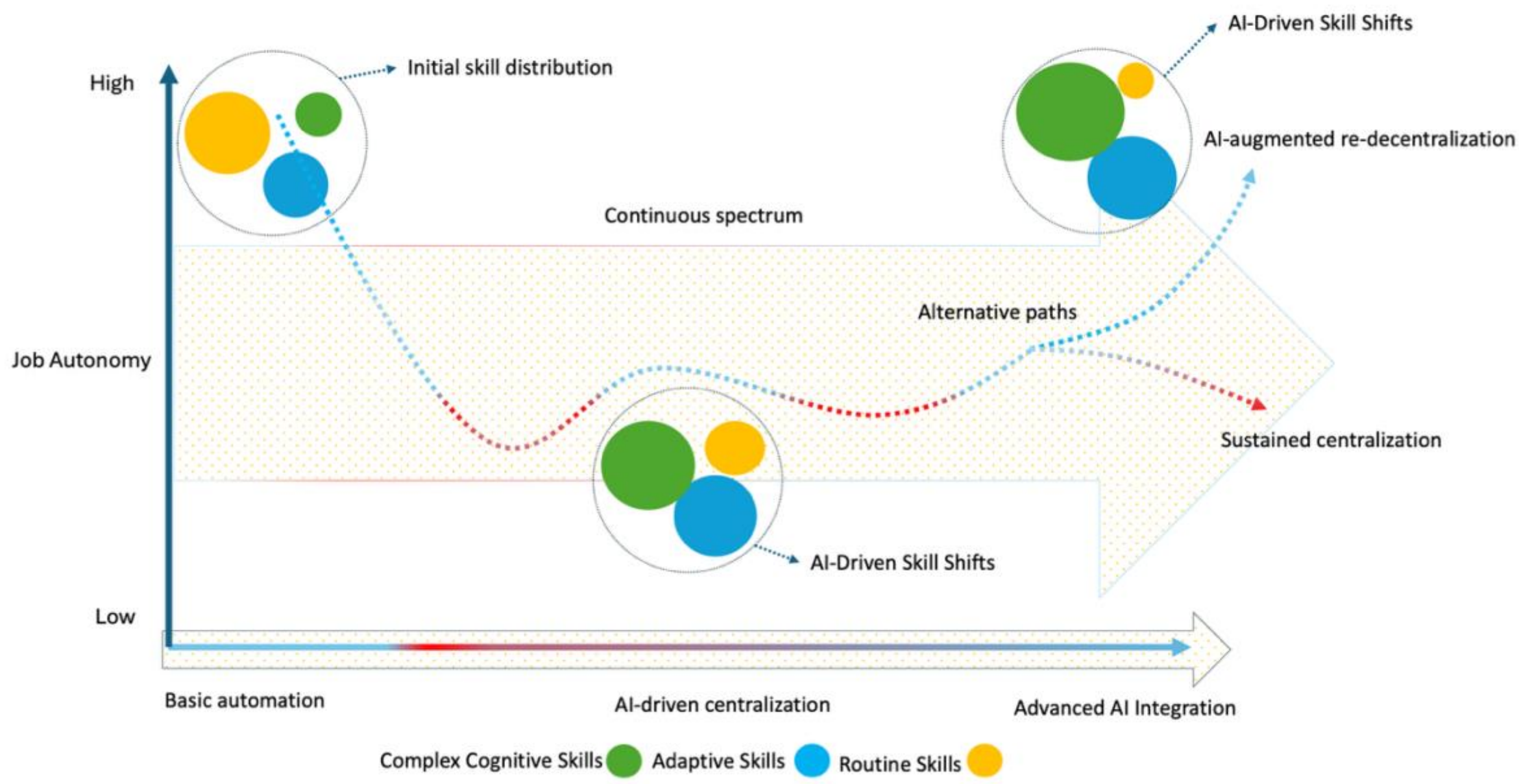

**FIGURE 1** *AI-Job Dynamics Framework*

For instance, in healthcare settings, Davenport and Kalakota (2019) observe that AI is increasingly used for administrative tasks such as scheduling and billing. This automation allows healthcare professionals to focus more on patient care, potentially expanding the range of medical skills they employ daily. While the AI handles routine administrative tasks, healthcare workers may engage in various patient-centric activities, from diagnosis to treatment planning and patient education.

## AI-Driven Centralization

As AI technologies mature and organizations accumulate more data, we observe a gradual shift, as organizations enter the second phase of the framework. As depicted in Figure 1, this middle phase shows increasing centralization through converging decision-making activities toward higher organizational levels. This phase is characterized by AI's growing capability to consolidate data processing and decision-making functions at an unprecedented scale. For instance, the trend in the rise of algorithmic management (Jarrahi et al., 2021), where AI-powered algorithms are increasingly used to oversee, direct, and evaluate workers, is driving the centralization of control within organizations. However, this transition is not abrupt or uniform. Most organizations are still in the early phases of AI integration, lacking extensive experience or resources for full implementation. This results in an uneven approach to centralization, creating a "wavy" path (Figure 1) where some organizations, departments, or functions embrace AI systems more rapidly than others. This heterogeneous progression sets the stage for the complex, non-linear changes in job characteristics we observe in this phase.

> *Proposition 4: In the AI-driven centralization phase, the relationships between AI-enabled data consolidation and decision-making and (a) job autonomy, (b) skill demands will be dynamic and non-linear, mediated by organizational learning processes.*

***Impact on job autonomy.*** During this phase, job autonomy often decreases as AI systems gradually assume decision-making roles across various organizational areas. This shift stems from AI's fundamental capability to centralize data processing and decision-making, enabling more efficient task performance than dispersed organizational units. Consequently, information and knowledge previously distributed across the organization become centralized, often concentrating decision-making power within specific departments or among select specialists (Günther et al., 2017). However, the extent of this centralization varies widely between organizations, suggesting a complex relationship between AI integration and job autonomy.

This variation across organizations is evident in empirical research. Meijer et al. (2021, p.844) demonstrate how largely the same AI technology deployed in different organizations (Berlin and Amsterdam police forces) can result in divergent outcomes driven by distinct organizational contexts and cultures. They found that in hierarchical and bureaucratic organizations, AI integration generated an "algorithmic cage," where workers experienced reduced autonomy and increased pressure to follow AI recommendations. In contrast, in organizations emphasizing professional judgment, AI became an "algorithmic colleague," serving as a tool to support rather than supplant human decision-making.

Furthermore, technological advancements have made large-scale data collection economically feasible, further driving centralization by incentivizing organizations to shift key decision-making processes to central bodies (Brynjolfsson & McAfee, 2014). This shift centralizes decision-making and impacts job autonomy at various organizational levels. As AI systems show their capability to process vast amounts of data and generate insights, the value traditionally placed on local expertise may gradually reduce. For instance, Brynjolfsson and Ng (2021) noted that a dataset at Walmart in Arkansas can provide insights into customer behaviors and predict purchasing patterns more accurately than individual grocery store managers across the US. A second example is the grocer ALDI, which uses AI to centralize its global shipping activities and optimize the entire supply chain. This eliminates the need for regional or local agents to manage their own logistics and supply chain decisions independently (ONE, 2024).

These examples highlight how AI-driven centralized systems can weaken localized decision-making over time, thereby reducing the autonomy of local managers. However, the impact on job autonomy may not be uniform across all contexts. The centralization trend may be less pronounced in organizations with strong cultures of local empowerment or in industries where tacit knowledge remains crucial. For instance, AI may augment rather than replace local decision-making in knowledge-intensive sectors, leading to a hybrid model of centralized data analysis and localized insight application (Faraj et al., 2018; Von Krogh, 2018).

This phase also aligns with observations on the potential for AI to centralize power among organizations (e.g., Acemoglu, 2021; Suleyman & Bhaskar, 2023). Large companies that have invested heavily in AI infrastructure and account for the majority of money spent on AI research are particularly prone to this centralization trend (Zuboff, 2019), though even they are still in the learning phase of effectively leveraging AI for decision-making across all organizational levels.

***Impact on skill variety.*** As AI systems gradually take over routine tasks and some decision-making functions, the demand for routine skills continues to reduce while the importance of adaptive and complex cognitive skills rises, as depicted in Figure 1. The circle representing routine skills in the figure reduces gradually, while those representing adaptive and complex cognitive skills expand markedly. This shift reflects the changing nature of human work in AI-augmented environments, aligning with Proposition 4b.

In this environment, workers increasingly need to develop new competencies related to AI interpretation and implementation and enhance their adaptive and complex problem-solving abilities to handle tasks that AI cannot easily automate. This transition period is marked by ongoing uncertainty and experimentation as organizations learn to navigate AI technologies' evolving possibilities and limitations. Organizations engage in both single-loop learning (improving existing processes) and double-loop learning (questioning fundamental assumptions and goals) (Argyris & Schön, 1978). This learning process requires workers to develop new skills in organizational adaptation and change management.

The heterogeneity in AI integration, stemming from the interplay of organizational learning processes and structural factors (Levitt & March, 1988), leads to varying impacts on skill variety across organizations and even departments. How organizations acquire, interpret, and act on new information and experiences (organizational learning processes) and their existing structures, hierarchies, and decision-making processes (structural factors) all influence AI integration. This varied integration pattern leads to diverse impacts on workforce skills. Some workers may need to rapidly upskill to work alongside advanced AI systems, while others may experience a more gradual shift in their skill requirements.

**Alternative Paths**

Building on the concept of AI endogenization introduced earlier, this section explores the co-evolutionary process between human agents and AI systems and its implications for organizational structures and practices. Figure 1 illustrates this advanced phase of AI integration where organizations face divergent paths - either maintaining centralization or moving toward re-decentralization, each with distinct implications for job autonomy and skill distributions. As organizations integrate AI systems, they adapt their understanding and practices, often leading to new, informal structures and behavior patterns that are not dictated by formal hierarchies. For instance, cross-functional teams augmented by AI capabilities may form organically to tackle complex problems, drawing on a rich combination of human expertise and AI-driven insights (Jansen et al., 2005).

While we focus on sustained centralization and AI-augmented re-decentralization as two significant alternative paths, it is important to note that these are not the only possible outcomes of AI endogenization. Organizations may experience combinations of these paths or develop entirely different structures as AI integration deepens. The following sections examine each path, examining their implications for organizational structure, job autonomy, and skill variety.

**Sustained Centralization**

In this scenario, as AI becomes deeply integrated into organizational processes, some organizations may maintain and reinforce centralized structures. As shown in the lower path of Figure 1, this trajectory is characterized by continued low levels of job autonomy and a distinct pattern of skill distribution. Although organizational adaptations and skill development occur, they are primarily directed toward optimizing centralized AI systems and decision-making processes. For instance, upskilling efforts might focus on enhancing employees' abilities to interact with and implement decisions from centralized AI systems rather than fostering decentralized decision-making capabilities. This approach reinforces existing centralized structures, with decision-making power remaining concentrated at higher organizational levels.

> *Proposition 5: In the sustained centralization phase, the degree of AI system effectiveness and data accumulation will be: (a) positively associated with further centralization of decision-making power; (b) negatively associated with job autonomy for most roles, except for those directly involved in AI management; (c) positively associated with skill polarization, increasing demand for specialized AI-related complex cognitive skills while reducing demand for routine skills.*

Several existing theories can help explain this tendency toward sustained centralization in AI-driven organizations, with significant implications for job autonomy and skill variety. W. Brian Arthur's theory of increasing returns (Arthur, 1989, 1994) explains how centralized AI systems become more effective as they accumulate data and computational power. This increased effectiveness attracts more users, generates more data, and further enhances the system's capabilities. This self-reinforcing cycle creates a "winner-takes-most" dynamic, where the benefits of centralization compound over time. Organizations that have benefited from the positive feedback loop and centralized AI capabilities are incentivized to maintain and reinforce these structures as the potential returns continue to increase. This dynamic can lead to sustained centralization, even as AI capabilities advance, potentially creating dominant players in various sectors and reshaping market structures.

Managerial incentives also play a role in driving sustained centralization. Brynjolfsson (2022) notes that managers often prefer task-level automation for its ease of implementation and measurement. Cost-cutting through automation is typically more straightforward than market expansion, and investors favor "scalable" business models that minimize labor intensity. In this context, managers might see no incentive to shift toward a more decentralized structure, even as AI becomes more deeply integrated into the organization.

***Impact on job autonomy and skill variety.*** As AI endogenization progresses toward sustained centralization, the concentration of data and decision-making power at the organizational core significantly impacts skill variety and job autonomy. We observe a continued trend toward specialized AI-related skills at the organizational core, while other roles may experience a reduction in skill variety. This aligns with the concept of skill-biased technological change (SBTC) discussed earlier (e.g., Acemoglu, 1998; Autor et al., 2003; Berman et al., 1998), illustrating how deep AI integration can reshape skill requirements. This polarization of skills creates a dichotomy within the workforce, where a small group of highly skilled AI specialists exists alongside a larger group of workers whose skills become increasingly standardized or automated as AI systems take over more complex tasks.

Simultaneously, job autonomy tends to decrease for many employees as decision-making becomes increasingly centralized and data-driven, reflecting the "digital Taylorism" described by Brown et al. (2010). Workers may find their decision-making capabilities constrained by algorithmic recommendations and restrictions, potentially leading to experiences of frustration, reduced autonomy, and job insecurity. These systems can also hinder effective communication between workers and managers (Kellogg et al., 2020).

However, while many roles may experience reductions in autonomy and skill variety, those involved in managing and interpreting AI systems may see increases in both, further contributing to workforce polarization.

## AI-Augmented Re-Decentralization

While sustained centralization represents one potential outcome of AI endogenization, the framework suggests that AI-augmented re-decentralization offers a viable alternative path. As depicted in the upper path of Figure 1, this trajectory shows an increase in job autonomy and a different configuration of skill requirements, particularly emphasizing complex cognitive and adaptive skills.

*Proposition 6: As the AI endogenization process deepens, the potential for AI-augmented re-decentralization increases, characterized by (a) a positive association with job autonomy across various organizational levels, (b) increased demand for complex cognitive and adaptive skills alongside a reduced demand for routine skills; and (c) a shift of decision-making power toward lower organizational levels, supported by AI.*

Several interrelated factors enable this path. A key technological driver is the rise of edge computing (Cao et al., 2020). As data generation increasingly occurs at the network's edge, it becomes more efficient to process data locally (Satyanarayanan, 2017; Shi & Dustdar, 2016). Edge computing supports real-time, context-aware decision-making at local levels, reducing the need for reliance on centralized systems (Shi et al., 2016). This shift is complemented by investments in AI infrastructure and initiatives aimed at reskilling and upskilling workers to collaborate effectively with AI systems (Davenport, 2018; Jaiswal et al., 2022).

Organizational culture also plays a crucial role in this transition. Entities with strong innovation cultures or decentralized decision-making traditions might more readily move toward this outcome (Klein & Sorra, 1996). Additionally, regulatory frameworks and organizational policies that encourage decentralized AI use and data governance can facilitate a shift toward re-decentralization.

Central to this process is a co-evolutionary dynamic where humans and intelligent machines mutually adapt and learn from each other (Raisch & Krakowski, 2021). This dynamic aligns with sociotechnical systems theory, emphasizing the reciprocal shaping of technological systems and social actors in organizations (Trist, 1981). In AI-augmented re-decentralization, this interaction fosters emergent organizational forms and practices, enabling more flexible and adaptive structures (Teece, 2016). As AI becomes more deeply integrated and workers become more adept at leveraging AI, decision-making power may shift back to lower organizational levels, though in a new, AI-augmented form. This represents a significant evolution in AI-driven organizational structures, combining the benefits of decentralized human judgment with the power of AI-driven insights and capabilities.

This co-evolutionary process highlights a key aspect of the framework: the reciprocal relationship between AI integration and job design elements. As organizations use AI systems to support decision-making at lower levels, job autonomy and skill requirements change. These changes, in turn, influence how AI is further integrated and utilized within the organization. For instance, as employees develop more sophisticated AI-related skills, they may discover new ways to leverage AI, potentially leading to greater decentralization of decision-making. This feedback loop between AI integration and job design elements drives the continuous evolution of organizational structures and practices, complementing the progression through AI integration phases discussed earlier.

***Impact on job autonomy***. In AI-augmented re-decentralization, job autonomy may increase across various organizational levels, reflecting a nuanced interplay between technological capabilities, human capital development, and organizational restructuring. This increase in autonomy stems from three key factors: advancements in AI technologies that enable local decision-making (including edge computing), the development of AI-related skills among workers, and the restructuring of organizations to accommodate decentralized decision-making.

Edge computing technologies enhance job autonomy by facilitating data processing closer to the data source, reducing latency, and supporting real-time, context-aware decision-making at local levels (Shi & Dustdar, 2016). Additionally, as workers develop more sophisticated AI-related skills through reskilling and upskilling initiatives, they gain a deeper understanding of AI capabilities and limitations. This enhanced comprehension allows them to make more informed choices about when and how to leverage AI insights, potentially increasing their discretion in decision-making processes.

The combination of advanced AI capabilities and more skilled workers creates opportunities for organizations to restructure, transferring decision-making authority to lower levels. For instance, in a retail context, this could empower store managers to make inventory decisions based on AI-processed local data, reducing reliance on centralized systems (Pan & McElhannon, 2018). This shift toward localized, AI-augmented decision-making represents a significant departure from centralized structures, potentially increasing job autonomy across various organizational levels.

***Impact on skill variety***. AI-augmented re-decentralization also significantly affects the skills required within organizations. As AI systems become more integrated, there is a growing demand for complex cognitive skills, such as technical competencies in AI and adaptive skills, while the need for routine skills decreases as AI systems automate many standardized tasks.

These skill shifts manifest uniquely in a decentralized context. For instance, in healthcare, local clinics might require staff who can interpret AI-generated diagnostic suggestions, adapt them for local patient needs, and communicate AI-driven decisions to patients (Topol, 2019). This decentralized approach demands technical AI skills but also strong communication and contextual adaptation abilities, reflecting the broader skill set needed when decision-making is decentralized (Lee et al., 2019; Von Krogh, 2018).

Unlike centralized models, where AI expertise is often concentrated at the top, re-decentralization requires a more widespread distribution of AI-related skills throughout the organization. This distribution allows local units to effectively use AI for context-specific decisions, increasing demand for a diverse skill set across organizational levels. As organizations progress, there may be a growing need for expertise in AI ethics, explainable AI (Barredo Arrieta et al., 2020; Floridi & Taddeo, 2016), and human-AI collaboration (e.g., Sharma et al., 2023), aligning with the framework’s emphasis on the dynamic nature of AI integration.

## DISCUSSION

The AI-Job Dynamics Framework presented in this paper opens several key avenues for future research and theoretical development in the field of AI and job design. By conceptualizing AI integration as a dynamic process with distinct phases, the framework encourages researchers to move beyond static, deterministic views of AI's impact on job characteristics.

To further develop and validate the framework, empirical validation through multi-level longitudinal studies is essential (Ployhart & Vandenberg, 2010). Such studies should track both organizations and workers as they navigate the AI integration process. At the organizational level, research should follow companies across various industries as they progress through different AI integration phases. For instance, examining how data-rich companies like Amazon or Walmart navigate these phases could provide valuable insights into the real-world dynamics of AI integration (Brynjolfsson et al., 2021). At the individual level, studies should track workers' experiences over time, similar to the Whitehall longitudinal research on job design and health (Marmot et al., 1991; Parker et al., 2017). These studies should use mixed-method approaches, combining quantitative measures of AI integration and job characteristics with qualitative assessments of worker experiences (e.g., Cordery et al., 2010; Jin & Rounds, 2012; Lesener et al., 2019).

Building on organizational learning processes outlined in Proposition 2, future research should explore how exploratory and exploitative learning (March, 1991) mediate the impact of AI integration on job characteristics. Studies could investigate how organizations balance these learning processes across different AI integration phases and how this balance influences job autonomy and skill variety outcomes.

The AI-Job Dynamics Framework also highlights AI-augmented re-decentralization as a potential outcome of AI endogenization, offering another rich area for investigation. Research could explore how advanced but less researched AI technologies, particularly edge computing, affect decision-making autonomy at various organizational levels. Key questions for future research include: Are our assumptions about how AI will challenge job characteristics correct? Are there other paths of technological advancement that we have yet to consider that may significantly impact job autonomy and skill variety? How might technologies like edge computing alter the trajectory of AI integration in ways not anticipated by the framework?

The framework, with its emphasis on non-linear processes and reciprocal causality, naturally gives rise to paradoxes and tensions in AI integration. Future studies should address these considerations, exploring questions such as: How do organizations manage the tension between centralized AI systems and localized human expertise, particularly in the AI-driven centralization phase? How does the simultaneous upskilling and deskilling effect of AI integration, as suggested in our discussion of skill variety, manifest across different organizational levels and sectors?

Several innovative methodological approaches could be used to empirically investigate these complex dynamics. Agent-based modeling (ABM) techniques offer a promising method for exploring the intricate interactions between AI systems, organizational structures, and individual workers that the framework posits (Davis et al., 2007; Gilbert, 2008). ABMs can simulate diverse, adaptive actors, allowing outcomes to emerge from their interactions rather than being predetermined. This approach is particularly suited to modeling the dynamic equilibria, cycles, or complex patterns that may arise in AI-augmented workplaces, aligning with the framework's emphasis on non-linearity and emergence (De Marchi & Page, 2014).

Additionally, natural language processing (NLP) can be used as a research tool to analyze corporate communications and employee feedback, offering a way to track changes in autonomy and skill requirements over time. By examining the linguistic patterns and sentiments within these data sources, researchers can gain insights into how these job characteristics evolve as organizations progress through different AI integration phases, as outlined in the framework (Hirschberg & Manning, 2015).

From a theoretical perspective, future research building on the AI-Job Dynamics Framework can extend or challenge existing organizational behavior and management theories. For instance, the traditional job characteristics model (Hackman & Oldham, 1976) was developed when automation primarily affected manual tasks, leaving cognitive jobs mainly in the human domain. However, with AI increasingly affecting cognitive work, a fundamental reconsideration of this model may be necessary. How might core job characteristics like skill variety, task identity, or autonomy be reconceptualized when AI systems can perform complex cognitive tasks, as outlined in the AI-driven centralization and AI-augmented re-decentralization phases?

Similarly, the framework's emphasis on organizational learning processes in AI integration raises questions about how organizational learning (Levitt & March, 1988) and knowledge management (Cyert & March, 1992) apply when knowledge is increasingly embedded in AI systems. How might these theories be extended to account for the reciprocal causality between AI integration and job design elements that the framework proposes?

## RESEARCH IMPLICATIONS

This study offers several important implications for research in AI and job design. First, by introducing the AI-Job Dynamics Framework, we provide a new lens to examine the evolving relationship between AI integration and key job characteristics. The framework challenges researchers to move beyond static, deterministic views of AI's impact on work, instead conceptualizing it as a dynamic, multi-directional process. This approach opens new opportunities for exploring how job autonomy and skill variety may shift non-linearly as organizations progress through different phases of AI integration.

We particularly emphasize the "alternative paths" concept in the AI endogenization process. We synthesize and extend existing scholarly work on augmentation and automation by proposing that the deep integration of AI can lead to divergent organizational outcomes, specifically identifying sustained centralization and AI-augmented re-decentralization as potential trajectories. This conceptualization provides a rich area for future empirical study. Researchers might explore how factors such as organizational culture, industry context, or technological capabilities influence an organization's path, thereby testing and refining our synthesis of previous theories in real-world contexts.

Finally, this study underscores the value of interdisciplinary research approaches to investigate the more salient effects of AI on jobs and organizational life more generally. The AI-Job Dynamics Framework draws on concepts from organizational behavior, information systems, and complexity theory. Bridging these disciplines—and incorporating insights from fields such as computer science and cognitive psychology—will be essential for a deeper understanding of how AI impacts job design and work more generally.